\documentclass[aps,prl,reprint,groupedaddress]{revtex4-1}

\pdfoutput=1
\usepackage[utf8]{inputenc}
\usepackage{amsmath,amssymb}
\usepackage{amsfonts}
\usepackage{slashed}
\usepackage{xcolor}

\usepackage[pdftex]{hyperref}
\hypersetup{ 
pdfstartview=FitV,
colorlinks=true, 
linkcolor=black, 
citecolor=blue, 
filecolor=black, 
urlcolor=magenta,
}
\usepackage{graphicx}

\newcommand{\inc}{\textrm{inc}}

\begin{document}


\title{DC resistivity of quantum critical, charge density wave states\\ from
    gauge-gravity duality}


\author{Andrea Amoretti}
\affiliation{Physique  Th\'{e}orique  et  Math\'{e}matique  and  International  Solvay  Institutes  Universit\'{e}  Libre de Bruxelles, C.P. 231, 1050 Brussels, Belgium}
\email{andrea.amoretti@ulb.ac.be}
\author{Daniel Are\'an}
\affiliation{Centro de F\'\i sica do Porto, Departamento de F\'\i sica da
Universidade do Porto, Rua do Campo Alegre 687, 4169-007 Porto, Portugal}
\email{daniel.arean@fc.up.pt}
\author{Blaise Gout\'eraux}
\affiliation{Nordita, KTH Royal Institute of Technology and Stockholm University, Roslagstullsbacken 23, SE-106 91 Stockholm, Sweden}
\email{blaise.gouteraux@su.se}
\author{and Daniele Musso}
\affiliation{Departamento  de  F\'{i}sica  de  Part\'{i}culas,  Universidade  de  Santiago  de  Compostela  and  Instituto  Galego  de  F\'{i}sica  de  Altas  Enerx\'{i}as  (IGFAE),  E-15782,  Santiago  de  Compostela, Spain.}
\email{daniele.musso@usc.es}

\date{\today}

\begin{abstract}
In contrast to metals with weak disorder, the resistivity of weakly-pinned charge density waves (CDWs) is not controlled by irrelevant processes relaxing momentum. Instead, the leading contribution is governed by incoherent, diffusive processes which do not drag momentum and can be evaluated in the clean limit. We compute analytically the dc resistivity for a family of holographic charge density wave quantum critical phases and discuss its temperature scaling. Depending on the critical exponents, the ground state can be conducting or insulating. We connect our results to dc electrical transport in underdoped cuprate high $T_c$ superconductors. We conclude by speculating on the possible relevance of unstable, semi-locally critical CDW states to the strange metallic region.
\end{abstract}


\maketitle

Part of the mystery shrouding the physics of high $T_c$ superconductors since their discovery over thirty years ago is tied to the nature of the ground state across the phase diagram \cite{Keimer2015}. There is an extremely rich pattern of both internal (superconductivity) and spacetime (nematicity, density waves, etc.)  symmetry breaking. There are strong indications that these orders can be intertwined \cite{2015RvMP...87..457F}, or even exist only as fluctuating phases \cite{RevModPhys.75.1201}. Understanding their impact on the low energy dynamics as well as their experimental signatures is challenging.

Spin-charge stripe order was originally detected at low temperatures $T<T_{CDW}$ in the pseudogap of underdoped lanthanum-based, neodymium-doped cuprates (La$_{2-x-y}$Nd$_y$Sr$_x$CuO$_4$ with $x\sim1/8$, $y\sim0.4$) \cite{Tranquada:1995}. Transport measurements \cite{PhysRevLett.88.147003} reveal finite frequency peaks in the far infrared regime compatible with weakly-pinned charge density waves (CDWs) \cite{RevModPhys.60.1129}. At $T<T_{CDW}$, the temperature dependence of the resistivity becomes insulating-like with a negative slope. It is also expected that this CDW order terminates at a zero temperature Quantum Critical Point (QCP) \cite{doi:10.1146/annurev-conmatphys-070909-104117}. More recently, experimental signatures of charge modulations on the overdoped side in a Bimuth-based compound (Bi2201) have also been reported \cite{2017arXiv170506165P}. At optimal doping, the strange metallic phase lies in the vicinity of the static CDW phase and is widely believed to originate from a QCP, although observables do not obey a simple scaling theory $\sigma(\omega,T)=T^{(d-2)/z}\Sigma(\omega/T)$ \cite{sachdev_2011}. This motivates studying quantum critical phases with spontaneously broken translations and more generally transport in strongly-coupled CDW phases.

Reliable theoretical tools to do so are few and far between. Field theoretical approaches (see \cite{2017arXiv170308172L} for a review and references therein) usually start by coupling a gapless, critical boson to a Fermi surface. In $d=2$ these theories are strongly-coupled in the IR and their analysis is intricate. Hydrodynamics or memory matrices can be used to capture transport properties at late times \cite{chaikin_lubensky_1995, RevModPhys.60.1129}. For static, weakly-pinned non-Galilean invariant CDW states, \cite{Delacretaz:2016ivq,Delacretaz:2017zxd} found that the resistivity is insensitive at leading order to irrelevant momentum-relaxing processes
\begin{equation}
\label{resweaklypinned}
\rho_{dc}=\frac1{\sigma_o}+O(\Gamma)\,,
\end{equation}
where $\Gamma$ is the momentum relaxation rate, and $\sigma_o$ is an incoherent conductivity. It represents the contribution to the conductivity of diffusive processes without momentum drag. This transport coefficient must be evaluated within the microscopic theory. It has been computed previously at a particle-hole symmetric $d=2$ spin density wave QCP \cite{Patel:2015dda}. Without particle-hole symmetry, distinguishing between the incoherent and momentum contributions is non-trivial \cite{Davison:2015bea}.

Gauge/Gravity duality \cite{Ammon:2015wua,Zaanen:2015oix,Hartnoll:2016apf} offers another approach by mapping the strongly-coupled dynamics to solving Einstein's equations in a weakly-coupled, classical theory of gravity. The incoherent conductivity can be computed analytically at non-zero density in both clean and disordered holographic metallic phases \cite{Davison:2015taa, Davison:2015bea}. Spatially modulated phases have been constructed using numerical methods \cite{Donos:2013wia,Withers:2013loa,Ling:2014saa,Jokela:2014dba,Jokela:2016xuy,Jokela:2017ltu,Andrade:2017cnc,Alberte:2017cch,Andrade:2017ghg}, which makes the study of the ground state and transport more challenging than when an analytical study is possible.

In this Letter, we combine Gauge/Gravity duality with Effective Field Theory (EFT) principles and investigate an effective holographic model of spontaneous translation symmetry breaking \cite{Amoretti:2016bxs}. We focus on the ground state and dc transport properties. The model breaks translations homogeneously \cite{Donos:2013eha,Andrade:2013gsa} rather than inhomogeneously, allowing us to obtain many results analytically. In a companion paper \cite{Amoretti:2017frz}, we explain in greater detail how it correctly captures various aspects of the EFT of CDW states \cite{chaikin_lubensky_1995, RevModPhys.60.1129} and give the technical derivations of our results. 

After briefly exposing the model, we address low frequency charge transport. We define the incoherent conductivity $\sigma_o$ and compute it analytically. We then construct hyperscaling violating holographic CDW QCPs and discuss the scaling of the incoherent conductivity at the QCP. Due to hyperscaling violation and the presence of irrelevant operators, $\sigma_o\neq T^{(d-2)/z}$ and it can be metallic or insulating. Finally, we connect our results to charge transport in cuprate high $T_c$ superconductors.

\section{Effective holographic theory of charge density waves}
We consider the family of holographic theories \footnote{All our considerations can be generalized to arbitrary dimension straightforwardly}:
\begin{equation}\label{action}
\begin{split}
&S=\int d^{4}x\,\sqrt{-g}\left[R-\frac12\partial\phi^2-V(\phi)\right.\\
&-\frac14\left(Z_1(\phi)+\lambda_1 Z_2(\phi)\sum_{I=1}^{2}\partial\psi_I^2\right)F^2\\
&\left.-\frac12\sum_{I=1}^{2}\left(Y_1(\phi)\partial \psi_I^2+\lambda_2 Y_2(\phi)\left(\partial \psi_I^2\right)^2\right)\right],
\end{split}
\end{equation}
where $I,J=1,2$ run over the boundary spatial coordinates. Our background Ansatz is \cite{Donos:2013eha,Andrade:2013gsa}:
\begin{equation}
\label{Ansatz}
\begin{split}
&ds^2=-D(r)dt^2+B(r)dr^2+C(r)d\vec x^2\,,\\
& A=A(r)dt\,,\quad \phi=\phi(r)\,,\quad \psi_I=k\delta_{Ii} x^i\,.
\end{split}
\end{equation}
The $\psi_I$'s break both spatial translations $x^i\to x^i+a^i$ and the global shift symmetry $\psi_I\to\psi_I+c_I$, but preserve a diagonal subgroup \cite{Nicolis:2013lma,Donos:2013eha,Andrade:2013gsa}, which is why the other bulk fields do not depend on the $x^i$ \footnote{Isotropy follows from a similar breaking of internal rotations of the $\psi_I$ and spacetime rotations down to a diagonal subgroup. As we have chosen the same value of $k$ in all spatial directions, the dual state preserves isotropy. Clearly this can be relaxed, with translations spontaneously broken anisotropically along one or several spatial directions.}. In what follows, we no longer distinguish between $i$ and $I$ indices.

We first restrict to the two-derivative action $\lambda_{1,2}=0$. The scalar couplings are arbitrary, except for their UV ($\phi\to0$) behavior $
V_{UV}=-6+\frac12m^2\phi^2+\dots,\quad Z_{1,UV}=1+\dots,\quad Y_{1,UV}=\phi^2+\dots$
which ensures asymptotically locally AdS$_{4}$ black holes exist when $r\to0$ \footnote{We have set the AdS radius to unity.}. The UV behavior of $Y_1(\phi)$ is one of the key points of our work: asymptotically, the three real scalars can be rewritten into two complex scalars $\Phi_I=\phi \exp(i\sqrt2\psi_I)/\sqrt2$, and the action becomes similar to that of Q-lattices \cite{Donos:2013eha}. This means we can think of the boundary theory as a UV CFT deformed by a pair of complex scalar operators $\mathcal L_{CFT}+(\lambda_I \mathcal O_I^\star+\lambda_I^\star\mathcal O^I)/2$ with dimension $\Delta$ given by $m^2=\Delta(\Delta-3)$. In a CDW state, the charge density is written as $\rho(x,t)=\rho_0+\rho_1(x,t)\cos(k x+\Psi(x,t))$ \cite{ RevModPhys.60.1129}. In the EFT, the order parameter is modeled by a complex scalar \cite{2015RvMP...87..457F}, the phase of which is expanded at linear order around equilibrium as $k x+\Psi(x,t)$. This is precisely what our boundary deformations capture \footnote{A similar model with $k=0$ was used in \cite{Iqbal:2010eh} to model holographic spin density waves. The properties of the corresponding Goldstones were further studied in \cite{Argurio:2015wgr}.}.

We do not expect the global shift symmetry to be an exact symmetry of the system at all energy scales. It is an emergent low energy symmetry tied to the dynamics of the Goldstones and is absent from inhomogeneous spatially modulated phases studied in past holographic literature. However, since we are concerned with low energy dynamics in this work, we treat this symmetry as an exact symmetry at all energy scales.

Imposing Dirichlet boundary conditions on $\phi$ with vanishing source $\phi\sim\phi_{(v)}r^\Delta+\dots$ together with $\psi_i=k x^i$ means translations are broken spontaneously \cite{Amoretti:2016bxs} rather than explicitly \cite{Donos:2013eha,Andrade:2013gsa,Donos:2014uba,Gouteraux:2014hca}.

The Lie derivative along $\partial/\partial_{\vec x}$ leaves all background fields invariant except the $\psi_i$ and generates a pure gauge solution to the equations of motion: this is the bulk dual to the boundary phonon \footnote{To get the dispersion relation, we need to include an $e^{-i\omega t+i q x}$ dependence, which will couple the $\psi_x$ perturbations to other fields.}. 

The higher-derivative couplings $\lambda_{1,2}$ source instabilities of translation invariant phases towards phases breaking translations spontaneously \cite{Amoretti:2017frz}. The free energy of the backreacted phases can be minimized with respect to the ordering wavevector $k$, thereby identifying the preferred $k\neq0$ of the ground state \cite{Donos:2013cka,Donos:2015eew}.

$\lambda_{1,2}$ are constrained by causality: \cite{Gouteraux:2016wxj} found a necessary condition on $\lambda_1$, $-1/6<\lambda_1<1/6$. We also take $\lambda_2>0$ and defer a more thorough analysis to future work. 

The dual, renormalized stress-energy tensor reads \cite{Amoretti:2017frz}:
\begin{equation}
\label{HoloST}
\langle T^{tt} \rangle= \epsilon=-3d_3=2\langle T^{ii} \rangle =2p\ ,
\end{equation}
where $\epsilon$ and $p$ are the energy density and the pressure, and $d_3$ appears in the boundary expansion of $r^2 D(r)\sim1+d_3 r^3+\ldots$ in the Fefferman-Graham gauge $B(r)=1/r^2$, together with the mimization condition
\begin{equation}
\label{ThermStabCond}
 k\int_{r_h}^0 \sqrt{BD}\,\left(Y_1 +2\lambda_2 k^2\frac{Y_2}C-\frac{\lambda_1Z_2A'^2}{BD}\right)\,=0\,.
\end{equation}
Restricting to a two derivative action in \eqref{action} $\lambda_1=\lambda_2=0$ would lead to $k=0$ for the ground state.

\eqref{HoloST} is compatible with the equilibrium stress tensor for a conformal Wigner crystal \cite{chaikin_lubensky_1995,Delacretaz:2017zxd} 
\begin{equation}
\label{eq:crystalST}
\langle T^{ij}_{eq}\rangle=\left[p+\left(K-G\right) \partial\cdot\langle\Psi\rangle\right]\delta^{ij}+2G\partial^{(i}\langle\Psi^{j)}\rangle
\end{equation}
 provided there is no phase gradient $\partial\cdot\langle \Psi\rangle$ at equilibrium. $K$ and $G$ are the bulk and shear moduli, which parameterize the elastic response of the phonons.

\section{Conductivity}
In a clean CDW state, translations are not broken explicitly. The low frequency electric conductivity reads:
\begin{equation}
\label{eq:cond}
\sigma(\omega)=\sigma_o+\frac{\rho^2}{\chi_{PP}}\frac{i}{\omega}\,.
\end{equation}
The imaginary $\omega=0$ pole comes from momentum conservation. At non-zero density $\rho$, the current overlaps with momentum and the dc conductivity is formally infinite. $\sigma_o$ is a finite, incoherent contribution to the real part. It appears as a transport coefficient in the constitutive relation of the electrical current $J^\mu=\rho u^\mu-T\sigma_o\partial^\mu(\mu/T)+\dots$ \cite{Delacretaz:2017zxd}, where $\mu$ the chemical potential, $T$ the temperature and $u^\mu$ the velocity. The incoherent conductivity is given by the Kubo formula
\begin{equation}
\label{eq:sigma0Kubo}
\sigma_o=\frac1{\left(\chi_{PP}\right)^2}\lim_{\omega\to0}\frac{i}{\omega}G^R_{J_{\inc}J_{\inc}}(\omega,q=0)\,.
\end{equation}
$J_\inc$ is the incoherent current orthogonal to momentum:
\begin{equation}
\label{JincDef}
J_{\inc} \equiv \chi_{PP} J-\rho P \,,\qquad \chi_{J_\inc P}=0\,.
\end{equation}
$\sigma_o$ is thus insensitive to momentum physics and can be expected to reflect universal properties of the QCP. It has been computed analytically in translation-invariant holographic phases \cite{Hartnoll:2007ip,Jain:2010ip,Chakrabarti:2010xy,Davison:2015taa}. In \cite{Amoretti:2017frz} we adapt an alternative computation described in \cite{Davison:2017}. Assuming the existence of a regular black hole horizon at $r=r_h$, we obtain
\begin{equation}
\label{eq:sigmarealpart}
\sigma_o=\left(\frac{s T}{sT+\mu\rho}\right)^2\left(Z_{1,h}+8\pi\lambda_1 k^2\frac{ Z_{2,h}}{s}\right),
\end{equation}
where $Z_{1,2,h}=Z_{1,2}(\phi(r_h))$ and we have used $\chi_{PP}=\epsilon+p=s T+\mu\rho$ \cite{Amoretti:2017frz}. In the translation-invariant limit $k\to0$, this matches previous literature \cite{Hartnoll:2007ip,Jain:2010ip,Chakrabarti:2010xy,Davison:2015taa}. When translations are spontaneously broken and for a thermodynamically preferred phase verifying \eqref{ThermStabCond}, it also agrees with the results of \cite{Donos:2018kkm}.

\section{DC resistivity of quantum critical CDWs}

Gauge/Gravity duality allows to model QCPs by constructing black hole solutions which display scaling behaviour in the radial coordinate in the deep IR region. The holographic radial coordinate can be thought of as a representation of the energy scale of the dual field theory, so that the UV of the field theory probes the region of spacetime close to the AdS boundary, while the IR probes the region close to the black hole horizon.

To model the QCPs, we truncate our holographic model \eqref{action} to its effective IR limit by taking \cite{Charmousis:2010zz,Gouteraux:2014hca}
\begin{equation}
\label{IRscalarcouplings}
V=V_0 e^{-\delta\phi},\quad Z_{i}=Z_{i,0} e^{\gamma_i\phi},\quad Y_{i}=Y_{i,0} e^{\nu_i\phi},
\end{equation}
with $i=1,2$.
This is the holographic equivalent of integrating out high energy modes to obtain the low energy effective action. We have assumed the scalar grows in the IR $\phi(\xi)=\kappa\log\xi\to\infty$, where $\xi$ is an IR radial coordinate, different from the $r$ coordinate which is defined over all of spacetime. The classical equations of motion of the IR effective action have the following solutions,  \cite{Amoretti:2017frz}:
\begin{equation}\label{skaska}
\begin{split}
& ds^2 = \xi^{\theta} \left[  \frac{L^2 d\xi^2}{\xi^2f(\xi) }-f(\xi)\frac{dt^2}{\xi^{2z}}  + \frac{d\vec{x}^2}{\xi^2}\right],\quad \psi_i=k x^i\,,\\
&f(\xi)=\left(1-\frac{\xi^{2+z-\theta}}{\xi_h^{2+z-\theta}}\right),\quad A = a\, \xi^{\zeta - z} dt\,.
 \end{split}
\end{equation} 
The IR is defined as $\xi\to+\infty$ ($\xi\to0$) when $\theta<2$ ($\theta>2$). The solution \eqref{skaska} is only valid in the near-horizon, IR region $\xi\gg 1$ ($\xi\ll1$) and indeed $\phi\to\infty$. The metric is parameterized by two exponents, $z$ and $\theta$. $z$ is the dynamical `Lifshitz' exponent: time and space scale anisotropically under rigid scale transformations $t\to\lambda^z t$, $\vec x\to\lambda \vec x$. Together with $\xi\to\lambda\xi$, the metric is seen to be covariant rather than invariant when $\theta\neq0$ \footnote{This means this scaling is not a symmetry, but a solution generating transformation.}. The temperature and entropy can be computed as usual from the surface gravity and the horizon area, $T\sim \xi_h^{-z}$, $s\sim T^{(2-\theta)/z}$: there is an effective dimensional crossover along the RG flow to $2-\theta$ spatial dimensions.

To connect to the UV, we need to perturb this IR solution with radial modes, which are analogous to irrelevant deformations of the IR endpoint of the RG flow. Some irrelevant modes decay sufficiently fast and decouple completely from the IR theory. However, others govern the leading behaviour of certain IR observables and can be \emph{dangerously irrevelant}. Whether modes are irrelevant or marginal is decided by how terms in the IR equations of motion coming from the Maxwell or $\psi_I$ matter sector scale with $\xi$: marginal (irrelevant) modes scale with the same (a subleading) power of $\xi$ as other terms. 

Our interest lies in determining the leading low temperature behaviour of $\sigma_o$ in the critical phases \eqref{skaska}.  \eqref{eq:sigmarealpart} reveals this is primarily determined by the low temperature scaling of $Z_1$. Indeed, if $\lambda_2$ sources a marginal coupling in the IR theory, then the second term inside the parentheses in \eqref{eq:sigmarealpart} has the $T$-dependence as the first. It will only affect the numerical prefactor and, barring very fine-tuned circumstances, will not cancel it out. If $\lambda_2$ sources an irrelevant deformation, then the second term inside the parentheses decays faster than the first as $T\to0$. Thus, the main lesson is that in both cases, it is enough to concentrate on the two-derivative part of the action to compute $\sigma_o(T\to0)$.

The IR solutions \eqref{skaska} were analyzed in some detail in \cite{Gouteraux:2014hca}. In general, $\kappa\gamma=2-\zeta$. 
$\zeta$ is related to the scaling dimension of one of the IR operators, $\Delta_{IR}=z+(2-\theta-\zeta)/2$. When $\zeta=\theta-2$, this deformation is marginal and the two-derivative Maxwell terms have the same $\xi$ scaling as other two-derivative terms in the eoms. If they decay faster in the IR, this deformation is irrelevant and gives the leading contribution to $\sigma_o$. Plugging this into \eqref{eq:sigmarealpart} leads to 
\begin{equation}
\label{eq:sigmarealpartlowT}
\sigma_o\sim T^{2+\frac{\zeta+2-2\theta}{z}}.
\end{equation}

As argued around \eqref{resweaklypinned}, $\sigma_o$ captures the dc resistivity of a weakly-pinned, quantum critical CDW state. In \cite{Delacretaz:2016ivq,Delacretaz:2017zxd}, it was shown that the ac conductivity is generally
\begin{equation}
\label{accondcdw}
\sigma(\omega)=\sigma_o+\left(\frac{\rho^2}{\chi_{PP}}\right)\frac{-i\omega}{\left(-i\omega(\Gamma-i\omega)+\omega_o^2\right)}\,,
\end{equation}
where $\Gamma$ and $\omega_o$ are the momentum relaxation rate and pinning parameter originating from weak explicit breaking of translations. The formula is valid when these parameters are small compared to the equilibration timescale $\Gamma,\omega_o\ll1/\tau_{eq}$.

Taking the dc limit $\omega\to0$ precisely returns \eqref{resweaklypinned}, so $\sigma_o$ captures the leading contribution up to $O(\Gamma,\omega_o)$ corrections. In contrast to weakly disordered metals where $\rho_{dc}\sim O(\Gamma)$, the resistivity is not small, even for weak disorder. Since weak disorder only affects subleading corrections, the resistivity can be computed in the clean limit using our formula \eqref{eq:sigmarealpart} or, at low temperatures, \eqref{eq:sigmarealpartlowT}.

Within the allowed parameter space, the zero temperature resistivity can diverge or vanish: these states can be insulating or conducting. This is a non-trivial feature of relaxing Galilean symmetry: in Galilean systems, $\sigma_o=0$ by symmetry and the CDW is always a dc insulator. The power can also vanish: then the resistivity saturates at zero temperature. This residual zero temperature resistivity is very reminiscent of the characteristics of charge transport in non-superconducting underdoped Nd-LSCO at $x\sim0.125$ with static spin-charge stripe order \cite{Tranquada:1995}. The resistivity is linear in temperature at $T>T_{CDW}$. At lower temperatures, it turns up and eventually asymptotes to a non-vanishing constant at zero temperature (see eg the inset of figure 2 of \cite{PhysRevLett.88.147003}).

At a QCP, a naive scaling analysis predicts $\sigma\sim T^{(d-2)/z}$ when $T$ is the only scale. Our result evades this expectation for two reasons: firstly, $[\sigma_o]=[\sigma_{inc}]-2[\chi_{PP}]$, where $\chi_{PP}$ contributes a dimensionful constant at $T=0$; secondly, $\theta\neq0$ and $\zeta\neq0$ also affect the $T$ dependence.

\section{Incoherent conductivity in unstable CDW critical phases}

We close this Letter by considering phases which break translations spontaneously but do not minimize the free energy. The motivation for this is as follows. \cite{Delacretaz:2016ivq} suggested that quantum critical CDW modulations provide a mechanism underlying room temperature off-axis peaks measured in the far infrared optical conductivity at optimal doping \cite{2004PMag...84.2847H}. If the strange metallic region is indeed the finite temperature wedge of a zero temperature QCP, any order parameter is subject to quantum critical fluctuations and there is no stable ordered phase.

Our main interest is the low temperature regime. Higher-derivative terms in \eqref{action} will either source subleading temperature dependence, or simply correct the prefactor of the two-derivative temperature scaling. So we set $\lambda_{1,2}=0$ in \eqref{action}. 

Since we do not impose the minimization condition \eqref{ThermStabCond}, matching the renormalized, dual stress-tensor to the crystal stress-tensor \eqref{eq:crystalST} now requires a uniform, non-zero strain $\partial\cdot\langle \Psi\rangle=\bar u$, \cite{Amoretti:2017frz}. The background is dual to an excited state which does not have the lowest free energy. It is similar to states with a non-zero superfluid velocity: these are generally less thermodynamically stable than equilibrium states without superfluid velocity.

The computation of the incoherent conductivity differs as now 
\begin{equation}
\chi_{PP}=\epsilon+p+2K\bar u\,,\quad K\bar u=-\frac{k^2}2\int_{r_h}^0 dr\sqrt{BD}Y_1
\end{equation}
which ultimately leads to
\begin{equation}
\label{eq:sigmarealpartunstable}
\sigma_o=\frac{\left(s T+2K\bar u\right)^2Z_{1,h}}{\left(\chi_{PP}\right)^2} +\frac{ 16\pi (K\bar u)^2 \rho^2}{s\,  Y_{1,h} k^2\left(\chi_{PP}\right)^2}.
\end{equation}

At low temperatures, neglecting the $sT$ terms in \eqref{eq:sigmarealpartunstable} and using $\epsilon+p\sim \mu\rho$, this becomes
\begin{equation}
\label{sigmainclowT}
\sigma_o\underset{T\to0}{\sim}\frac{\left(2K\bar u\right)^2}{\left(\mu\rho+2K\bar u\right)^2} \left(Z_h+\frac{ 4\pi \rho^2}{s\,  Y_h k^2}\right).
\end{equation}
An interesting subcase is the so-called semi-locally critical limit, where both $z\to+\infty$ and $\theta\to-\infty$, with $-\theta/z=1$. In phases breaking translations explicitly, this leads to a $T$-linear heat capacity and resistivity \cite{Davison:2013txa}. Its connexion with Planckian dissipation and potential relevance to optimally doped cuprates has been highlighted in \cite{Davison:2013txa}. Importantly, \cite{Davison:2013txa} studied a case with only marginal deformations, so the only IR scale is $T$ and $\tau_{eq}\sim\tau_{P}=\hbar/k_B T$. \cite{Davison:2013txa} also focused on the slow momentum-relaxing regime, where momentum relaxation dominates the resistivity $\rho_{dc}\sim\Gamma\sim1/\tau_P$ and the incoherent conductivity is negligible.

\eqref{eq:sigmarealpartunstable} applies to phases breaking translations spontaneously rather than explicitly, and shows that in the semi-locally critical limit, the incoherent conductivity also provides a $T$-linear, Planckian contribution to the resistivity when there are only marginal deformations. For this, it is crucial that the state does not mimize the free energy: the same limit taken in \eqref{eq:sigmarealpartlowT} leads to $\sigma_o\sim T^3$.

\cite{Delacretaz:2016ivq} determined the dc conductivity of weakly-pinned, fluctuating CDWs to be
\begin{equation}
\sigma_{dc}=\sigma_o+\frac{\rho^2}{\chi_{PP}}\frac{\Omega}{\Omega\Gamma+\omega_o^2}
\end{equation}
where $\Omega$ is the phase relaxation rate of the phonons. In a quantum critical phase, it should be thought of as setting the timescale of phase fluctuations. The data catalogued in \cite{Delacretaz:2016ivq} suggests the Planckian timescale $\tau_{P}$ controls both the dc and ac transport in bad metals, including the strange metal region of cuprates. It would be interesting to compute holographically this second contribution to the resistivity.

\paragraph{Note added:} As this work was in the final stages, we became aware of \cite{Alberte:2017oqx} which also studies a homogeneous model of spontaneous translation symmetry breaking in holographic massive gravity.

\paragraph{Second note added:} After this work appeared as a preprint, \cite{Donos:2018kkm} emphasized how considering thermodynamically stable phases affects the incoherent conductivity. This new version reflects this improved understanding.

\begin{acknowledgments}
We would like to thank Riccardo Argurio for collaboration at an early stage. We would like to thank Matteo Baggioli, Carlos Hoyos, Francisco Ib\'{a}\~{n}ez, Elias Kiritsis, Sasha Krikun, Nicodemo Magnoli, Alfonso Ramallo, Javier Tarr\'\i o, Paolo di Vecchia and Jan Zaanen for stimulating and insightful discussions. We are grateful to Sean Hartnoll for comments on a previous version of the manuscript. BG has been partially supported during this work by the Marie Curie International Outgoing Fellowship nr 624054 within the 7th European Community Framework Programme FP7/2007-2013. The work of D.M. was supported by grants FPA2014-52218-P from Ministerio de Econom\'\i a y Competitividad. D.A. is supported by the 7th Framework Programme (Marie Curie Actions) under grant agreement 317089 (GATIS) from the grant CERN/FIS-NUC/0045/2015, and by the Simons Foundation grants 488637 and 488649 (Simons collaboration on the Non-perturbative bootstrap). D.A. and D.M. thank the FRont Of pro-Galician Scientists for unconditional support. B.G. would like to thank the institute AstroParticle and Cosmology, Paris for warm hospitality at various stages of this work.
\end{acknowledgments}

\bibliography{STSB}
\end{document}